\documentclass[aps,prd,twocolumn,showpacs,amssymb,refmerge,superscriptaddress,floatfix]{revtex4-1}
\usepackage{longtable}
\usepackage{amsmath}
\usepackage{graphicx} 
\usepackage{dcolumn}  
\usepackage{color}
\usepackage{subfigure}
\usepackage[colorlinks,linkcolor=blue,urlcolor=blue,anchorcolor=blue,citecolor=blue]{hyperref}
\usepackage{bm}

\begin{document}

\title{Light-front holographic QCD with generic dilaton profile}

\newcommand*{\PKU}{School of Physics and State Key Laboratory of Nuclear Physics and Technology, Peking University, Beijing 100871,China}\affiliation{\PKU}
\newcommand*{\CICQ}{Collaborative Innovation Center of Quantum Matter, Beijing, China}\affiliation{\CICQ}
\newcommand*{\DUKE}{Department of Physics, Duke Univeristy, Durham, North Carolina 27708, USA}\affiliation{\DUKE}
\newcommand*{\CHEP}{Center for High Energy Physics, Peking University, Beijing 100871, China}\affiliation{\CHEP}

\author{Zhaoheng Guo}\email{zhaohengguo@pku.edu.cn}\affiliation{\PKU}
\author{Tianbo Liu}\email{liutb@jlab.org}\affiliation{\PKU}\affiliation{\DUKE}
\author{Bo-Qiang Ma}\email{mabq@pku.edu.cn}\affiliation{\PKU}\affiliation{\CICQ}\affiliation{\CHEP}

\begin{abstract}
\noindent

We generalize the soft-wall and hard-wall models to a light-front holographic QCD model with a generic dilaton profile. The effective potential induced by a higher power dilaton profile is interpreted as a stronger color confinement at long distance, and it gradually evolves to the hard-wall model when the power increases to infinity. As an application, we investigate the exotic meson states recently discovered in experiments in the generic soft-wall model with a higher power dilaton profile, and the results are in agreement with the spectra of the exotic mesons. Our calculation indicates a weaker interaction at short distance and a stronger interaction at large distance for the components in the exotic mesons. The generic dilaton profile deserves further scrutiny for understanding the strong interaction and for applications.

\end{abstract}


\maketitle

QCD provides a fundamental description of the strong interaction in terms of quark and gluon degrees of freedom in the standard model. The asymptotic freedom property of QCD allows perturbative calculation to be validly applied. However, the nonperturbative and nonlinear nature of QCD at the low energy scale makes it a challenging task to directly derive all hadronic properties from first principles. During the last decade, the light-front holographic approach~\cite{Brodsky:2003px}, which is oriented from the light-front quantization~\cite{Brodsky:1997de} and the correspondence between the gravity in the anti-de Sitter (AdS) spacetime and the conformal field theory (CFT)~\cite{Maldacena:1997re} in the physical spacetime, has proven successful in explaining many hadronic properties~\cite{Brodsky:2014yha} and thus has become an alternative nonperturbative method to study hadron physics. Although the conformal symmetry of the classical QCD Lagrangian with massless quarks is broken by the quantum effects, its asymptotic freedom at high energy and the infrared (IR) fixed point inspired by lattice simulation~\cite{Furui:2006py}, the Dyson-Schwinger equation~\cite{vonSmekal:1997is}, and the empirical effective charge~\cite{Deur:2005cf} indicate that QCD is nearly conformal at some region. Recently, a remarkable relation between the light-front dynamics and the conformal quantum mechanics has also been established~\cite{Brodsky:2013ar} through the de Alfaro-Fubini-Furlan (dAFF) mechanism~\cite{deAlfaro:1976je}. Therefore, the AdS/CFT correspondence can be used as a tool to obtain the first approximation to QCD.

In the light-front holographic framework, the fifth-dimensional coordinate $z$ in the AdS space is exactly mapped to the Lorentz invariant variable $\zeta$, which measures the separation of the constituents in the hadron~\cite{Brodsky:2007hb}. As a direct result of the confinement, an IR cutoff should be imposed on the coordinate $z$. This cutoff is also understood as an energy scale, which breaks the conformal symmetry and induces the discrete spectra in QCD. At present, there are two methods, referred to as the hard-wall model~\cite{Polchinski:2001tt} and the soft-wall model~\cite{Karch:2006pv}, to introduce the IR cutoff. In the former method, a sharp cutoff is imposed at large distance, and one considers a slice of the AdS space $0\leq z\leq z_0$ with some boundary conditions at the IR border. In the latter method, a dilaton background is introduced as a deformation of the AdS space and effectively a smooth cutoff at large distance, but its profile is largely unspecified. Practically, a quadratic form is usually adopted. It can reproduce the Regge trajectory~\cite{Karch:2006pv,Brodsky:2008pg,Abidin:2009hr} and the massless pion in the chiral limit~\cite{Brodsky:2013ar}.

In this paper, we prove that the hard-wall model can be viewed as the limit of the soft-wall model with the power of the dilaton profile tending to infinity. Then, as an application, we investigate the exotic mesons in the generic soft-wall model with a dilaton beyond the quadratic form, and the results are in good agreement with the experimental data of the exotic meson spectra. Therefore, the soft-wall model with a generic dilaton background is not only of theoretical interest but also of physical significance to understand the strong interaction.

Hadrons are eigenstates of the QCD light-front Hamiltonian $H_\textrm{LF}=P_{\mu}P^{\mu}=2P^+P^-  - \bm{P}_\perp^2$ with the invariant mass square as the eigenvalues,
\begin{equation}
H_{\textrm{LF}}\bigl\vert \psi_{\textrm{H}} \bigr\rangle= M^2_{\textrm{H}}\bigl\vert \psi_{\textrm{H}} \bigr\rangle.
\end{equation}
Quantized at fixed light-front time $\tau=(t+z)/\sqrt{2}$, a hadron state can be expanded on the Fock state bases as
\begin{equation}
\bigl\vert \psi_{\textrm{H}} \bigr\rangle
= \sum_{n}
  \int \left[d x\right]  \left[d^2 \bm{k}_{\perp}\right]  \psi_{n/\textrm{H}}(x_i,\bm{k}_{\perp i})
\bigl\vert n: x_i, \bm{k}_{\perp i}, \lambda^z_{i} \bigr\rangle,
\end{equation}
where the integral measures are
\begin{align}
  \int \left[d x\right] &= \prod_{i=1}^n \int dx_i \delta(1-\sum_{j=1}^nx_j),\\
  \int \left[d^2 \bm{k}_{\perp}\right] &= \prod_{i=1}^n \int \frac{ d^2 \bm{k}_{\perp}}{2(2\pi)^3} 16\pi^3\delta^{(2)}(\sum_{j=1}^n\bm{k}_{\perp j}).
\end{align}
The variables $x$ and $\bm{k}_\perp$ are respectively the light-front momentum fraction and the intrinsic transverse momentum carried by the component. The coefficients $\psi_{n/\textrm{H}}(x_i,\bm{k}_{\perp i})$ are the light-front wave functions (LFWFs), which are frame independent. They encode all partonic information in the hadron. Then the mass square of the hadron can be expressed in terms of the LFWFs as
\begin{align}
M_{\textrm{H}}^2  &=  \int \big[d x\big]  \left[d^2 \bm{k}_{\perp}\right]
 \sum_{i=1}^n \left(\frac{ \bm{k}_{\perp i}^2 + m_i^2 }{x_i} \right)
 \left\vert \psi (x_i, \bm{k}_{\perp i}) \right \vert^2 \nonumber \\
  &\qquad + \psi^*(x_i,\bm{k}_{\perp i})U\psi(x_i,\bm{k}_{\perp i}),\label{eqmass}
\end{align}
where $U$ is an effective potential. With a Fourier transformation, Eq.~\eqref{eqmass} can be expressed in the coordinate space, which is more convenient to investigate the potential. One main purpose of the light-front holographic approach is to obtain the effective potential.

As proved in the conformal limit~\cite{Brodsky:2007hb}, the fifth-dimensional coordinate $z$ in the AdS space exactly corresponds to a Lorentz invariant light-front variable $\zeta$, which measures the separation of the constituents in the hadron. For a two-body system, it is expressed as
\begin{equation}
  \zeta^2=x(1-x)\bm{b}_\perp^2,
\end{equation}
where $\bm{b}_\perp$ is the Fourier conjugate of the intrinsic momentum $\bm{k}_\perp$. Then, the eigenequation is expressed as
\begin{equation}
\left[-\frac{d^2}{d\zeta^2}-\frac{1-4L^2}{4\zeta^2}+U(\zeta)\right]\phi(\zeta)=M_{\textrm{H}}^2\phi(\zeta),
\label{gequation}
\end{equation}
where the masses of the constituents are neglected. The $\phi(\zeta)$ is the transverse mode of the LFWF:
\begin{equation}
  \psi(x, \zeta, \theta) = e^{iL\theta}X(x)\frac{\phi(\zeta)}{\sqrt{2\pi\zeta}}.
\end{equation}

Now the only remaining issue is to determine the effective potential $U(\zeta)$, which describes the confinement. In the hard-wall model, a sharp cutoff is applied,
\begin{align}
  U(\zeta)=\left\{
  \begin{matrix}
    0 & (\zeta < \zeta_0)\\
    \infty & (\zeta > \zeta_0)
  \end{matrix}\right..
\end{align}
It can be viewed as a relativistic version of the MIT bag model. From Eq.~\eqref{gequation}, one may find that the hadron mass spectra are given by the roots of the Bessel function $M_{L,n}=\beta_{L,n}/\zeta_0$~\cite{Brodsky:2006uqa}. It has an asymptotic linear behavior $M\sim 2n+L$.

\begin{figure}
\centering
\includegraphics[width=0.35\textwidth]{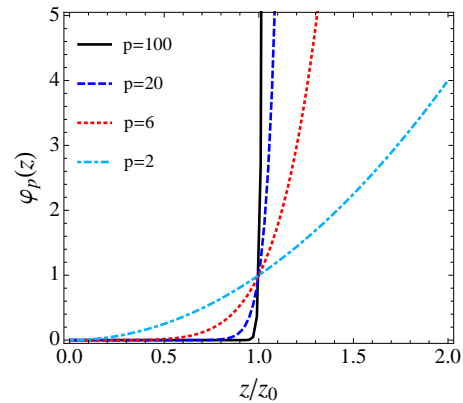}
\caption{The generic dilaton profile with different power $p$.\label{fdilaton}}
\end{figure}

\begin{figure}
\centering
\includegraphics[width=0.35\textwidth]{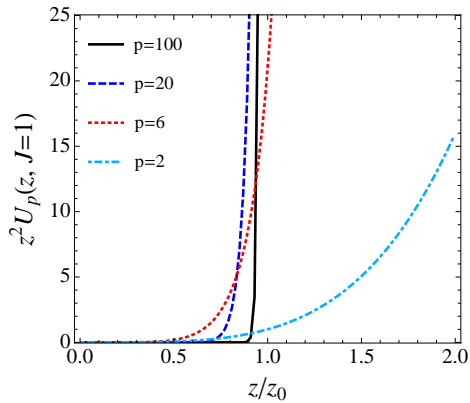}
\caption{The effective potentials induced by different dilaton profiles.\label{fpotential}}
\end{figure}

In the soft-wall model, a dilaton background $\varphi(z)$ is introduced as a deformation of the AdS space. It explicitly breaks the conformal symmetry and induces an effective potential,
\begin{equation}
U(z, J)=\frac{1}{2}\varphi^{\prime \prime} +\frac{1}{4}{\varphi^{\prime}}^2+\frac{2J-3}{2z}\varphi^{\prime},
\label{effpotential}
\end{equation}
where $J$ is the spin of the hadron. In practice, a quadratic dilaton profile $\varphi(z)=\kappa^2z^2$ is usually adopted due to its success in phenomenology. However, the dilaton in the soft-wall approach is viewed as a smooth cutoff in the IR limit. Therefore, it should reduce to the hard-wall situation at some limit. For this purpose, a generic dilaton profile should be introduced. For simplicity but still universal, we choose a dilaton profile as
\begin{equation}
\varphi_p(z)=\kappa^pz^p.
\end{equation}
In Fig.~\ref{fdilaton}, we plot the dilaton profile with different $p$. Then, the effective potential is correspondingly expressed as
\begin{align}
  U_p(z,J)&=\frac{p^2+2(J-2)p}{2}\kappa^pz^{p-2}+\frac{p^2}{4}\kappa^{2p}z^{2p-2}\nonumber\\
  &=\frac{p^2+2(J-2)p}{2}\kappa^2\left(\frac{z}{z_0}\right)^{p-2}+\frac{p^2}{4}\kappa^2\left(\frac{z}{z_0}\right)^{2p-2},
  \label{genepotential}
\end{align}
where $z_0$ is a typical length scale defined as $z_0=1/\kappa$. One may observe from Fig.~\ref{fpotential} that the potential in the short distance region $z<z_0$ decreases to zero with respect to an increasing value of $p$, while in the large distance region $z>z_0$ it gradually increases to infinity. Thus, it is natural to expect that the generic soft-wall model reduces to the hard-wall model, when $p$ goes to infinity.

In the limit of $p\to\infty$,
\begin{equation}
\lim_{p\to\infty}e^{-\kappa^{p}z^{p}}=\theta(1-\kappa^2 z^2),
\end{equation}
where $\theta(x)$ is the Heaviside step function with the convention $\theta(0)=1/e$. Correspondingly, a positive dilaton is expressed as
\begin{align}
e^{\varphi_\infty(z)}=\left\{
\begin{array}{lc}
1 & (z<z_0)\\
e^{-1} & (z=z_0)\\
\lim_{\epsilon\to0}\epsilon^{-1} & (z>z_0)
\end{array}
\right..
\end{align}
Then, the confinement potential is
\begin{align}
U_\infty(z,J)=\left\{
\begin{array}{lc}
0 & (z<z_0)\\
\infty & (z \geq z_0)
\end{array}\right.,
\end{align}
which means an infinite high potential in the long distance region and no interaction in the short distance region. This is exactly the picture of the hard-wall model. The value at $z=z_0$ does not affect the solutions of the differential equation, since it is a null set.

One may also view this procedure from the aspect of the eigenequation. Substituting the effective potential \eqref{genepotential} into the Eq.~\eqref{gequation}, we rewrite the eigenequation as
\begin{equation}\label{mequation}
  \left[-\frac{d^2}{d\tilde{z}^2}-\frac{1-4L^2}{4\tilde{z}^2}+U_p(\tilde{z},J)\right]\phi(\tilde{z})=z_0^2M_{\textrm{H}}^2\phi(\tilde{z}),
\end{equation}
where $\tilde{z}=z/z_0$. Obviously, the invariant mass square is proportional to $\kappa^2$. In Table \ref{eigenvalue}, we list the eigenvalues $M_n^2/\kappa^2$ evaluated with different choices of $p$. One can find that the separation between the adjacent eigenvalues increases with the value of $p$. This is consistent with the Wentzel-Kramers-Brillouin (WKB) semiclassical approximation. In the limit of $p\to\infty$, the eigenvalues converge to those calculated from the hard-wall approach with a cutoff at $z_0$. Simultaneously, the eigenfunctions converge to the Bessel form $\phi(\tilde{z})\sim\sqrt{\tilde{z}}J_L(\beta_{L,n}\tilde{z})$, which is the solution in the hard-wall model. Thus, we prove that the soft-wall model with the generic dilaton background reduces to the hard-wall situation, when the power $p$ in the profile tends to infinity. In other words, both the hard-wall and the soft-wall approaches are unified into a generalized soft-wall model.

\begin{table}
  \centering
  \caption{The eigenvalues in the soft-wall model with different choices of $p$. The spin $J$ is assigned to $1$.\label{eigenvalue}}
\begin{tabular}{c c c c c c}
\hline
\hline
p  & $M_1^2/\kappa^2$ & $M_2^2/\kappa^2$ & $M_3^2/\kappa^2$ & $M_4^2/\kappa^2$ &  $M_5^2/\kappa^2$ \\
\hline
2  & 2 & 6 & 10 & 14 & 18 \\
3  & 4.00 & 13.85 & 26.17 & 40.21 & 55.63 \\
6  & 5.89 & 26.05 & 56.16 & 94.86 & 141.28 \\
10 & 6.14 & 30.30 & 69.59 & 122.35 & 187.84 \\
20 & 6.06 & 31.51 & 76.09 & 138.61 & 218.30 \\
80 & 5.86 & 30.89 & 75.88 & 140.79 & 225.57 \\
400 & 5.65 & 29.77 & 73.17 & 135.84 & 217.80 \\
Hard-wall & 5.77 & 30.47 & 74.89 & 139.04 & 222.93 \\
\hline\hline
\end{tabular}
\end{table}

\begin{figure}
\centering
\includegraphics[width=0.35\textwidth]{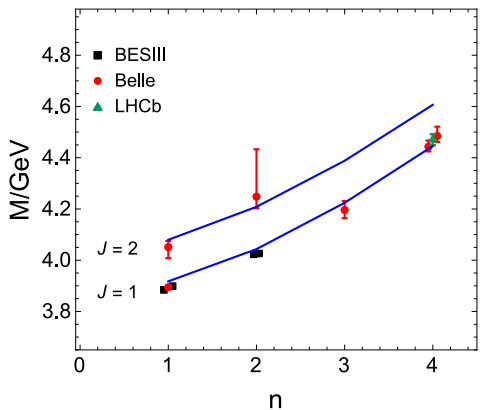}
\caption{The mass spectra of the $Z$ mesons. The solid curve is the model results with $\kappa=223$\,MeV. The data are taken from
Refs.~\cite{Ablikim:2013mio,Ablikim:2015tbp,Ablikim:2013xfr,Ablikim:2013wzq,Ablikim:2013emm,Liu:2013dau,Chilikin:2014bkk,Mizuk:2009da,Chilikin:2013tch,Mizuk:2008me,Aaij:2014jqa}.}
\label{spectrum_figure}
\end{figure}

As an application, we investigate the recently discovered charmoniumlike charged meson states, often referred to as $Z$ states~\cite{Ablikim:2013mio,Ablikim:2015tbp,Ablikim:2013xfr,Ablikim:2013wzq,Ablikim:2013emm,Liu:2013dau,Chilikin:2014bkk,Mizuk:2009da,Chilikin:2013tch,Mizuk:2008me,Aaij:2014jqa}, together with their isospin partners, in the generalized soft-wall model with a dilaton beyond the quadratic form. We choose the dilaton profile with $p=6$, which corresponds to an effective potential
\begin{equation}
  U_6(\tilde{z},J)=6(J+1)\kappa^2\tilde{z}^4+\kappa^2\tilde{z}^{10}.
\end{equation}
Then, the invariant mass squares are solved from Eq.~\eqref{mequation}. We should note that the charmoniumlike $Z$ meson states are usually interpreted as the tetraquark states, which means their lowest Fock states contain four quarks, i.e., $|c\bar{c}q\bar{q}\rangle$, where $q$ represents the light quark. In this case, we cannot simply neglect the mass of the constituents, as the charm quark mass is comparable to a hadron mass. Thus, a mass shift $\Delta m$ is required. At present, there are many phenomenological models aiming to unravel the structure of these states, such as the molecule model~\cite{Tornqvist:1991ks}, the hadro-quarkonium model~\cite{Dubynskiy:2008mq}, and the diquark-antidiquark model~\cite{Maiani:2004uc,Brodsky:2014xia}. In our calculation, the meson states are regarded as two-body systems, which means that the tetraquark state is divided into two clusters. Here, we treat each cluster as a two-quark system. Then, the mass shift is obtained from the mass of the cluster. We adopt the mass values of the $c\bar{q}$ system calculated in the soft-wall model~\cite{Branz:2010ub} as $m=1.86$\,GeV for a scalar state and $m=2.02$\,GeV for a vector state. In our scheme, $Z_c$(3900)~\cite{Ablikim:2013mio, Ablikim:2015tbp, Ablikim:2013xfr, Liu:2013dau}, $Z_c$(4020)~\cite{Ablikim:2013wzq,Ablikim:2013emm}, $Z_c$(4200)~\cite{Chilikin:2014bkk}, and $Z$(4430)~\cite{Mizuk:2009da,Chilikin:2013tch,Aaij:2014jqa} are categorized into a series of radial excitations with $J = 1$, while $Z_1$(4050) and $Z_2$(4250)~\cite{Mizuk:2008me} are arranged into the excitations with $J = 2$. In Table~\ref{spectrum_table}, we compare the theoretical calculations with the data of the spectra. In Fig.~\ref{spectrum_figure}, we show the tendency of the mass spectra with respect to the radial quantum number $n$. Our results are in good agreement with the experimental data. Therefore, the generalized soft-wall model with a dilaton beyond the quadratic form is not only of theoretical interest to unify the soft-wall and the hard-wall models but also of physical significance for phenomenological applications. This will enrich our understanding of the strong interaction.

\begin{table}
\caption{The mass spectra of the $Z$ mesons.\label{spectrum_table}}
\begin{tabular*}{0.5\textwidth}{m{0.12\textwidth}m{0.12\textwidth}m{0.2\textwidth}r}
\hline\hline
 Exotic mesons  & $M_{\textrm{th}}$~(MeV) & $M_{\textrm{ex}}$~(MeV) & Ref.\\
\hline
$Z_c(3900)$ & $3917.5$ & $3899.0\pm3.6\pm4.9$ & \cite{Ablikim:2013mio} \\
$Z_c(3900)$ & $3917.5$ & $3894.5\pm6.6\pm4.5$ & \cite{Liu:2013dau} \\
$Z_c(3900)$ & $3917.5$ & $3894.8\pm2.4\pm3.2$ & \cite{Ablikim:2015tbp} \\
$Z_c(3900)$ & $3917.5$ & $3883.9\pm1.5\pm4.2$ & \cite{Ablikim:2013xfr} \\
$Z_c(4020)$ & $4043.3$ & $4022.9\pm0.8\pm2.7$ & \cite{Ablikim:2013wzq} \\
$Z_c(4020)$ & $4043.3$ & $4026.3\pm2.6\pm3.7$ & \cite{Ablikim:2013emm} \\
$Z_c(4200)$ & $4224.1$ & $4196^{+31+17}_{-29-13}$ & \cite{Chilikin:2014bkk} \\
$Z(4430)$ & $4445.8$ & $4443^{+15+19}_{-12-13}$ & \cite{Mizuk:2009da} \\
$Z(4430)$ & $4445.8$ & $4485^{+22+28}_{-22-11}$ & \cite{Chilikin:2013tch} \\
$Z(4430)$ & $4445.8$ & $4475\pm7^{+15}_{-25}$ & \cite{Aaij:2014jqa} \\
$Z_1(4050)$ & $4079.7$ & $4051\pm14^{+20}_{-41}$ & \cite{Mizuk:2008me}\\
$Z_2(4250)$ & $4208.4$ & $4248^{+44+180}_{-29-35}$ & \cite{Mizuk:2008me}\\
\hline\hline
\end{tabular*}
\end{table}

We should emphasize that our study does not refute the remarkable success of the soft-wall model with the quadratic dilaton profile. The choice of the quadratic form can reproduce the Regge trajectory and the massless pion in the chiral limit. It is also supported by the dAFF mechanism and the superconformal algebra~\cite{deTeramond:2014asa}. Thus, the effective potential induced by the quadratic dilaton profile provides a good description of the interaction between a quark and an antiquark or that between an active quark and a cluster carrying the quantum numbers of an effective spectator.

However, the conformal property of QCD is valid in the limit of zero quark mass. The finite quark mass inevitably breaks the strictly conformal symmetry, but it is important to reproduce the baryon properties, such as the spin structure~\cite{Liu:2015jna} for the realistic hadron world. In this sense, some kind of generalization of the quadratic dilaton profile seems to be optional in some specific situations for describing the strong interaction in the AdS/QCD framework. Furthermore, the exotic mesons or the tetraquark states do not obey the Regge trajectory; nor are they viewed as Goldstone bosons in the chiral perturbative theory. Thus, these phenomenological results are no longer constraints on the choice of the dilaton profile when one investigates the exotic states. On the other hand, the effective potential induced by a higher power dilaton profile is interpreted as the interaction between two clusters, each containing two quarks inside. Hence, the effective potentials utilized here do not conflict with the usual one, as they describe different situations.

We also point out that the confinement potential from a higher power dilaton leads to a weaker interaction in the
short distance region but a stronger interaction in the large distance region. This provides a physical picture that the
clusters feel relatively weak interaction in the hadron, but when the separation increases, they meet a much stronger
confinement. Thus, the generic dilaton profile meets the QCD basic properties of asymptotic freedom at short distance
and color confinement at large distance.

In summary, we generalize the light-front holographic soft-wall model with a generic dilaton profile. We prove that the generalized soft-wall model reduces to the hard-wall model in the limit of an infinite power. In other words, the soft-wall and hard-wall approaches are unified into one general model. As an application, we investigate the exotic meson states in the generalized soft-wall model with a higher power form of the dilaton background. The calculated results are consistent with the experimental data of exotic meson spectra.

\acknowledgments{
This work is supported by the National Natural Science Foundation of China (Grants No.~11035003, No.~11120101004 and No.~11475006). It is also supported by the Principal Fund for Undergraduate Research at Peking University, the Training Program of Innovation for Undergraduates of Beijing, the National Fund for Fostering Talents of Basic
Science (Grants No.~J1103205 and No.~J1103206), and the US Department of Energy under contract numbers DE-FG02-03ER41231.}

\end{document}